\documentclass{emulateapj}

\shorttitle{Two Supernovae in IC 883}
\shortauthors{Kankare et al.}

\def\kms{\ifmmode{\rm km\,s^{-1}}\else\hbox{$\rm km\,s^{-1}$}\fi}

\begin{document}


\title{Discovery of Two Supernovae in the Nuclear Regions of the Luminous Infrared Galaxy IC 883}


\author{E. Kankare\altaffilmark{1}, S. Mattila\altaffilmark{1}, S. Ryder\altaffilmark{2}, P. V\"ais\"anen\altaffilmark{3}, A. Alberdi\altaffilmark{4}, A. Alonso-Herrero\altaffilmark{5,6}, L. Colina\altaffilmark{5},\\ 
A. Efstathiou\altaffilmark{7}, J. Kotilainen\altaffilmark{8}, J. Melinder\altaffilmark{9}, M.-A. P\'erez-Torres\altaffilmark{4}, C. Romero-Ca\~nizales\altaffilmark{4}, A. Takalo\altaffilmark{1}}
\altaffiltext{1}{Tuorla Observatory, Department of Physics and Astronomy, University of Turku, V\"ais\"al\"antie 20, FI-21500 Piikki\"o, Finland \\ erkki.kankare@utu.fi}
\altaffiltext{2}{Australian Astronomical Observatory, PO Box 296, Epping, NSW 1710, Australia.}
\altaffiltext{3}{South African Astronomical Observatory, PO Box 9, Observatory 7935, South Africa.}
\altaffiltext{4}{Instituto de Astrofísica de Andalucia, IAA-CSIC, Apartado 3004, 18080 Granada, Spain}
\altaffiltext{5}{Departamento de Astrof\'isica, Centro de Astrobiolog\'ia, CSIC/INTA, Carretera de Torrej\'on a Ajalvir, km 4, 28850, Torrej\'on de Ardoz, Madrid, Spain}
\altaffiltext{6}{Instituto de F\'{\i}sica de Cantabria, CSIC-Universidad de Cantabria, 39005 Santander, Spain}
\altaffiltext{7}{School of Sciences, European University Cyprus, Diogenes Street, Engomi, 1516 Nicosia, Cyprus}
\altaffiltext{8}{Finnish Centre for Astronomy with ESO (FINCA), University of Turku, V\"ais\"al\"antie 20, FI-21500 Piikki\"o, Finland}
\altaffiltext{9}{Department of Astronomy, Oskar Klein Centre, Stockholm University, AlbaNova University Centre, 106 91 Stockholm, Sweden}




\begin{abstract}
We report the discovery of two consecutive supernovae (SNe), \object{2010cu} and \object{2011hi}, located at 0.37\arcsec\ (180 pc) and 0.79\arcsec\ (380 pc) projected distance respectively from the centre of the \textit{K}-band nucleus of the luminous infrared galaxy \object{IC 883}. The SNe were discovered in an ongoing near-infrared \textit{K}-band search for core-collapse SNe in such galaxies using the ALTAIR/NIRI adaptive optics system with laser guide star at the Gemini-North Telescope. These are thus the closest SNe yet discovered to a LIRG nucleus in optical or near-infrared wavelengths. The near-infrared light curves and colours of both SNe are consistent with core-collapse events. Both SNe seem to suffer from relatively low host galaxy extinction suggesting that regardless of their low projected galactocentric distances, they are not deeply buried in the nuclear regions of the host galaxy.
\end{abstract}

\section{Introduction}

Luminous ($L_{{\rm IR}} > 10^{11} L_{\odot}$) and ultraluminous ($L_{{\rm IR}} > 10^{12} L_{\odot}$) infrared (IR) galaxies (LIRGs and ULIRGs, respectively), have high star formation (SF) rates. The fraction of SF in U/LIRGs in the local universe is small compared to that in normal spiral galaxies; however, at high redshift they become the dominant sources of SF \citep[e.g.,][]{magnelli11}. Stars more massive than $\sim8M_{\sun}$ explode as core-collapse supernovae (CCSNe), and due to their short life cycles they are a very useful tool for tracing ongoing SF rates, independent of the conventional way of using the galaxy IR luminosity. Recently, \citet{anderson11} have presented evidence that the SN population in the interacting LIRG system \object{Arp 299} differs from those observed in normal spiral galaxies with the Ib and IIb SNe being more numerous and centrally concentrated than more common Type II SNe. Until now, this kind of study has been impossible due to insufficient numbers of SNe discovered in U/LIRGs, providing a strong motivation for SN searches in galaxies with high SF rates. The extinction-free searches at radio wavelengths using very long baseline interferometry (VLBI) have been successful in discovering SNe and SN remnants in U/LIRGs, e.g., in \object{Arp 220} \citep[][]{parra07, batejat11} and \object{Arp 299} \citep{perez-torres09, ulvestad09}. The reason for the low optical discovery rate of CCSNe in U/LIRGs is the high dust extinction combined with spatial concentration of SF in the crowded nuclear regions within the central few hundreds of parsecs \citep[see e.g.,][]{soifer01}. Therefore imaging in the near-IR where the extinction is strongly reduced, and high spatial resolution is achievable with ground-based adaptive optics (AO) or space-based imaging is required. Recently such methods have been shown to be very efficient with the discovery of several SNe in the LIRG nuclear/circumnuclear regions. \object{SN 2004ip} was discovered with the Very Large Telescope using the natural guide star NACO AO system \citep{mattila07}. \object{SNe 2008cs} and \object{2004iq} reported in \citet{kankare08} were discovered using the Gemini-North Telescope with the laser guide star (LGS) assisted ALTAIR AO system and the \textit{Hubble Space Telescope} (\textit{HST}) NICMOS archive data, respectively.  

In this Letter, we report the consecutive discovery of two SNe in the same host galaxy with the Gemini-North Telescope\footnote{Based on observations obtained at the Gemini Observatory, which is operated by the Association of Universities for Research in Astronomy, Inc., under a cooperative agreement with the NSF on behalf of the Gemini partnership: the National Science Foundation (United States), the Science and Technology Facilities Council (United Kingdom), the National Research Council (Canada), CONICYT (Chile), the Australian Research Council (Australia), Minist\'{e}rio da Ci\^{e}ncia e Tecnologia (Brazil) and Ministerio de Ciencia, Tecnolog\'{i}a e Innovaci\'{o}n Productiva (Argentina).}, \object{SN 2010cu} and \object{SN 2011hi}. \citet{sanders03} report the host LIRG \object{IC 883} (also known as e.g., \object{UGC 8387}, \object{Arp 193}, and \object{IRAS 13183+3423}) to have an IR luminosity of $L_{IR} = L[8-1000 \mu m] = 4.7 \times 10^{11} L_{\odot}$ and a distance of 100 Mpc (H$_{0}$ = 75 km~s$^{-1}$Mpc$^{-1}$) giving a projected linear distance scale $1\arcsec = 485$~pc. The IR luminosity of \object{IC 883} indicates a CCSN rate of $\sim1.3$ yr$^{-1}$ derived with the empirical relation of \citet{mattila01}, consistent with the discovery of two SNe within two years. In \citet{romero-canizales11b} we report the details of our radio follow-up of \object{SNe 2010cu} and \object{2011hi} and investigate the innermost nuclear regions of \object{IC 883} through high-angular resolution, high-sensitivity radio observations. 

\section{Observations and Results}

\subsection{Near-IR Observations}

\object{SNe 2010cu} and \object{2011hi} were discovered as a result of our \textit{K}-band search for highly-obscured CCSNe in a sample of eight nearby LIRGs using the Gemini-North Telescope with the Near-InfraRed Imager (NIRI) with ALTAIR LGS AO system (0.022\arcsec\,pixel$^{-1}$, FWHM $\sim$ 0.1\arcsec). The individual NIRI images were reduced with {\sc iraf} using the NIRI package and registered and co-added using a bright field star as a reference. The discovery image of \object{SN 2010cu} was observed on 2010 February 24.6 UT (program GN-2010A-Q-40, PI: S. Ryder). Comparison to a previously obtained ALTAIR/NIRI \textit{K}-band reference image from 2009 June 10.4 UT was done using a slightly modified version of the image subtraction package ISIS 2.2 \citep{alard98,alard00}. The software matches the point-spread functions (PSFs) and flux levels of a previously aligned pair of images by deriving an optimal convolution kernel. The image subtraction process revealed a new point source located only 0.37\arcsec\ from the brightest \textit{K}-band core of \object{IC 883} corresponding to a projected distance of $\sim$180 pc. The transient was reported to the Central Bureau for Astronomical Telegrams (CBAT) immediately after the discovery and was designated as a possible supernova \object{PSN K1002-1} \citep{kankare10}. A follow-up observation in \textit{JHK} on 2010 May 4.5 UT confirmed its SN nature resulting in a formal SN designation \object{SN 2010cu} \citep{ryder10}. Further follow-up observations were obtained on 2010 June 4.3 UT in \textit{HK} and on 2011 February 11.6 UT in \textit{JHK} (GN-2011A-Q-48, PI: S. Ryder). As expected close to a year from discovery \object{SN 2010cu} had faded below the detection limit. However, a new point source had appeared 0.79\arcsec\ (projected distance $\sim$380 pc) from the \textit{K}-band nucleus of the host galaxy. It was reported as a possible supernova \citep{kankare11a} and follow-up observations in \textit{JHK} were obtained on 2011 April 18.5 UT, leading to a formal designation \object{SN 2011hi} \citep{kankare11b}. The \textit{K}-band discovery images of \object{SNe 2010cu} and \object{2011hi} and a subtraction between these two are shown in Fig.~1. We note that the projected distance of \object{SN 2010cu} places it along the dusty ring reported by \citet{clemens04}, whereas \object{SN 2011hi} is located outside the ring.

\begin{figure*}[t]
\epsscale{1.10}
\plotone{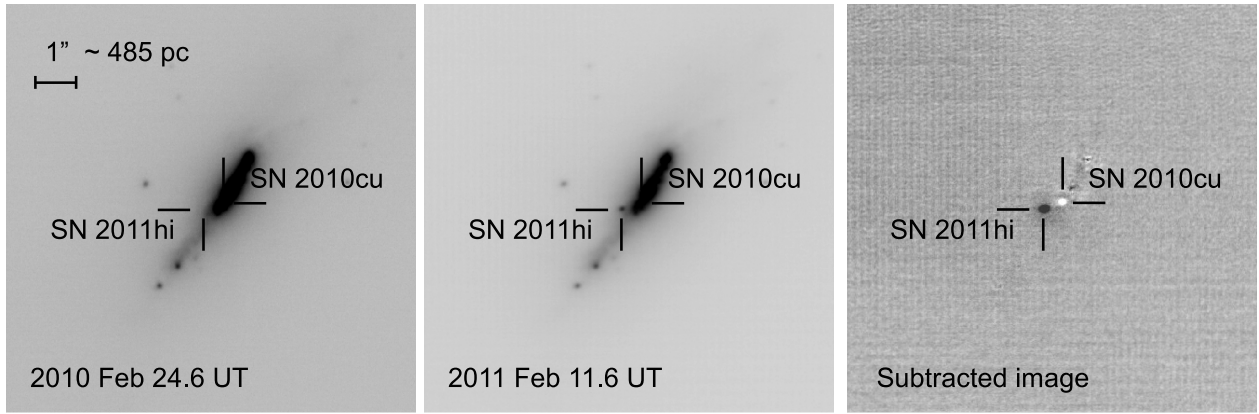}
\caption{10\arcsec$\times$10\arcsec\ subsections of \textit{K}-band Gemini ALTAIR/NIRI AO discovery images of SNe 2010cu (left panel) and 2011hi (middle panel) and the subtraction between these two (right panel). The smooth subtraction of the host galaxy IC 883 clearly demonstrates the good alignment and PSF match between the images and the two SNe can be clearly detected as individual point sources thanks to the high angular resolution provided by the AO. North is up and East is to the left.}
\end{figure*}

Follow-up near-IR observations of \object{SNe 2010cu} and \object{2011hi} were also attempted with the Nordic Optical Telescope\footnote{Based on observations made with the Nordic Optical Telescope, operated on the island of La Palma jointly by Denmark, Finland, Iceland, Norway, and Sweden, in the Spanish Observatorio del Roque de los Muchachos of the Instituto de Astrofisica de Canarias} (NOT) using the NOTCam instrument (0.234\arcsec\,pixel$^{-1}$) on 2010 March 8.3 UT (seeing FWHM $\sim$ 0.7\arcsec) in \textit{HKs} and 2011 February 20.2 UT (FWHM $\sim$ 0.9\arcsec) in \textit{JHKs}. However, neither epoch provided detection of the SNe even with the help of image subtraction with the reference images obtained with the same instrument on 2009 March 15.1 UT (FWHM $\sim$ 1.1\arcsec). This demonstrates the importance of high spatial resolution for SN searches in the nuclear regions of LIRGs. 

Due to the small field-of-view (FOV) of the ALTAIR/NIRI AO setup of only 22.5\arcsec\ $\times$ 22.5\arcsec\ an accurate World Coordinate System (WCS) was derived with an iterative method. First the WCS was derived for a 2MASS \textit{J}-band image of the field using 47 isolated point sources from the 2MASS catalogue with astrometric accuracy of 0.1\arcsec\ \citep{skrutskie06}. The 2MASS image was then aligned to an \textit{HST}/ACS \textit{F814W} image of the galaxy with a 3.5\arcmin\ $\times$ 3.5\arcmin\ FOV (obtained from the \textit{HST} Science Archive) using 8 sources common to both images. Finally the \textit{HST} image was transformed to the Gemini images using 14 point-like sources. Both transformation processes allowed shifts in x and y, scaling and rotating. This yielded R.A. $= 13^{\mathrm{h}}20^{\mathrm{m}}35.354^{\mathrm{s}}$ and Decl. $= +34^{\circ}08\arcmin21.86\arcsec$ for \object{SN 2010cu} and R.A. $= 13^{\mathrm{h}}20^{\mathrm{m}}35.387^{\mathrm{s}}$ and Decl. $= +34^{\circ}08\arcmin21.69\arcsec$ for \object{SN 2011hi}. For the brightest \textit{K}-band core of \object{IC 883} we obtain R.A. $=13^{\mathrm{h}}20^{\mathrm{m}}35.336^{\mathrm{s}}$ and Decl. $= +34^{\circ}08\arcmin22.16\arcsec$. The rms errors for the derived coordinates are 0.28\arcsec\ and 0.21\arcsec\ for R.A. and Decl., respectively. The alignment of the 2MASS image to match the \textit{HST} image dominates the errors. The \textit{K}-band core of \object{IC 883} is within 1$\sigma$ from the 6.9~GHz radio centre reported in \citet{romero-canizales11b}.

\subsection{Photometry and Extinctions}

The Gemini images were calibrated with an isolated field star in the FOV, \object{13203564+3408148} in the 2MASS catalogue, using the standard star \object{FS 131} \citep{hawarden01} observations in \textit{JHK} obtained on 2011 April 18.5 UT. Both stars were measured with aperture photometry using GAIA and aperture-corrected to a radius of 100pix (=2.2\arcsec). This yielded $m_{J}=17.44 \pm 0.02$~mag, $m_{H}=16.77 \pm 0.02$~mag and $m_{K}=16.50 \pm 0.02$~mag for the field star with the errors being dominated by the statistical photometric error. The photometry of \object{SNe 2010cu} and \object{2011hi} was derived with the QUBA\footnote{Python package specifically designed for imaging and spectroscopy reduction and SN photometry in the {\sc iraf} environment developed by S. Valenti (Queen’s University, Belfast).} pipeline using the {\sc snoopy}\footnote{SNOOPY, originally presented in Patat (1996), has been implemented in {\sc iraf} by E. Cappellaro. The package is based on daophot, but optimised for SN magnitude measurements.} package, both running standard {\sc iraf} tasks. Three point sources in the ALTAIR/NIRI FOV were used to derive the PSF over the images used for fitting the SN in the subtracted images. Photometric errors were estimated by subtracting the fitted SN from the image and measuring several artificial sources simulated around the SN position with a dispersion of 1~$\times$ FWHM. This error usually dominated over the statistical uncertainty of the PSF fitting. The \textit{J}-band 5$\sigma$ upper limit for \object{SN 2010cu} was derived based on the faintest detected sources in the image taking into account the increased noise due to the image subtraction process. Table~1 lists apparent SN magnitudes, where errors are in parentheses. 

\begin{table}
\caption{Photometry of SNe 2010cu and 2011hi.}
\centering
\begin{tabular}{cccc}
JD & $m_{J}$ & $m_{H}$ & $m_{K}$\\
(2400000+) & (mag) & (mag) & (mag) \\ 
\hline
SN 2010cu & & & \\
\hline
55252.1 & - & -	& 17.36(0.06) \\
55321.0	& $>$19.7 & 19.18(0.09) & 18.74(0.15) \\
55351.8	& - & 19.52(0.05) & 19.24(0.09) \\
\hline
SN 2011hi & & & \\
\hline
55604.1 & 17.92(0.08) & 17.20(0.05) & 16.67(0.04) \\
55670.0 & 18.07(0.06) & 17.00(0.05) & 16.42(0.04) \\
\hline
\end{tabular}
\end{table}

The NOT images were also used for deriving the \textit{JHK} magnitudes of the ALTAIR/NIRI field star for a consistency check. Again, the QUBA pipeline was used and the following magnitudes obtained for the 2009 reference epoch, $m_{J}=17.39 \pm 0.11$~mag, $m_{H}=16.81 \pm 0.09$~mag, $m_{K}=16.46 \pm 0.18$~mag using seven 2MASS stars for the calibration.

As a simple approach to estimate the SN type and the host galaxy extinction in the lines-of-sight to \object{SNe 2010cu} and \object{2011hi} the \textit{JHK} light curve templates from \citet{mattila01} were used. These templates are divided into two subtypes denoted 'ordinary' (linearly declining) and 'slowly declining' (based on Type IIL SN 1979C and Type IIn SN 1998S). The 'ordinary' template contains a variety of SNe including Types Ib/c, IIL and even IIb and IIP; despite some intrinsic differences in their near-IR light curves, we consider the template to describe at least the Type Ib/c SNe fairly well. The 'slowly declining' template shows a near-IR excess originating from pre-existing dust in the circumstellar medium (CSM) of the progenitor \citep{fassia00}. The well sampled near-IR light curves for the prototypical Type IIP \object{SN 1999em} from \citet{krisciunas09} were also used to represent the class of Type II SNe showing a plateau phase. From \citet{schlegel98} we obtain a Galactic extinction of A$_{V}=0.041$ for the \object{IC 883} line-of-sight. For the host galaxy extinction both \citet{cardelli89} and \citet{calzetti00} reddening laws were considered and only the former for the Galactic extinction. We used $\chi^{2}$-fitting to derive the best fit for \textit{V}-band host galaxy extinction A$_{V}$, the discovery epoch $t_{0}$ and the same constant shift $C$ for the absolute magnitude in all the three bands. For the 'ordinary' and 'slowly declining' templates $t_{0}$ is given relative to the \textit{K}-band peak and for Type IIP with respect to the explosion date. The magnitude shift $C$ represents the intrinsic luminosity differences between the SNe and the templates. The \textit{J}-band upper limit of \object{SN 2010cu} was not included in the fit. Results and reduced $\chi^{2}$ values are reported in Table~2 and the fits for each SN are shown in Fig.~2.

\begin{table*}
\caption{Best $\chi^{2}$ fits for the SN parameters.}
\centering
\begin{tabular}{c | cccc | cccc}
Template & $A_{V}$ & $t_{0}$ & $C$ & $\tilde{\chi}^{2}$ & $A_{V}$ & $t_{0}$ & $C$ & $\tilde{\chi}^{2}$\\
 & (mag) & (days) & (mag) & & (mag) & (days) & (mag) & \\ 
\hline
SN 2010cu & \multicolumn{4}{|c|}{Cardelli law} & \multicolumn{4}{|c}{Calzetti law} \\
\hline
ordinary & 0.1 & -8 & +0.80 & 11 & 0.1 & -8 & +0.80 & 11\\
slow & 0.0 & 6 & +2.20 & 8.5 & 0.0 & 6 & +2.20 & 8.5\\
IIP & 1.3 & 62 & +0.10 & 1.7 & 1.0 & 62 & +0.15 & 1.7\\
\hline
SN 2011hi & \multicolumn{4}{|c|}{ } & \multicolumn{4}{|c}{ } \\
\hline
ordinary & 6.8 & -38 & -1.35 & 5.0 & 4.7 & -38 & -1.00 & 4.4\\
slow & 0.0 & 89 & -0.25 & 27 & 0.0 & 89 & -0.25 & 27\\
IIP & 7.0 & 31 & -1.65 & 9.4 & 5.0 & 31 & -1.30 & 8.8\\
\hline
\multicolumn{9}{c}{ }\\
\multicolumn{9}{c}{ }\\
\end{tabular}
\end{table*}

\begin{figure*}
\epsscale{1.00}
\plotone{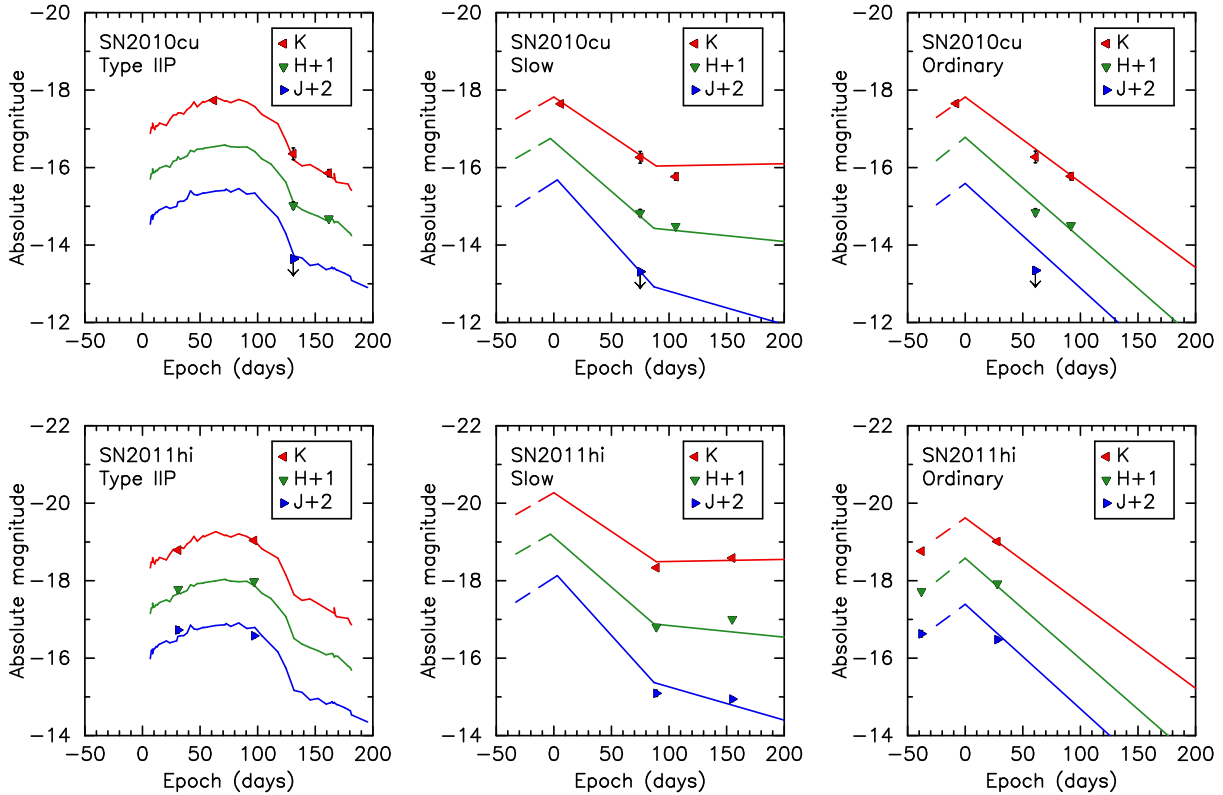}
\caption{Template light curve fits for SN 2010cu and SN 2011hi are shown. The SN absolute magnitudes have been corrected for the derived total line-of-sight extinctions adopting the Calzetti extinction law. The template curves have been shifted in magnitude relative to the data by a constant $C$. The epoch for 'slowly declining' and 'ordinary' fits is relative to the \textit{K}-band peak and for the Type IIP to the explosion date.}
\end{figure*}

From SN-free F110W and F222M images observed in a \textit{HST}/NICMOS survey of LIRGs by \citet{scoville00} and obtained from the \textit{HST} archive we produced a $J-K$ colour map of the central regions of \object{IC 883}, see Fig.~3. The NICMOS filter magnitudes were converted into Vega magnitudes using the transformations of \citet{origlia00}. Assuming a typical intrinsic evolved stellar population colour of $J-K\sim0.9$~mag \citep[e.g.,][]{fioc99} and the Calzetti extinction law the observed near-IR colour of the galaxy can be converted into total line-of-sight extinction below the limit $\tau_{J}\sim1.0$ i.e., with extinctions higher than $A_{V}\sim3$~mag only the surface extinction is mapped and thus the values represent a lower limit. 

\begin{figure}
\epsscale{1.00}
\plotone{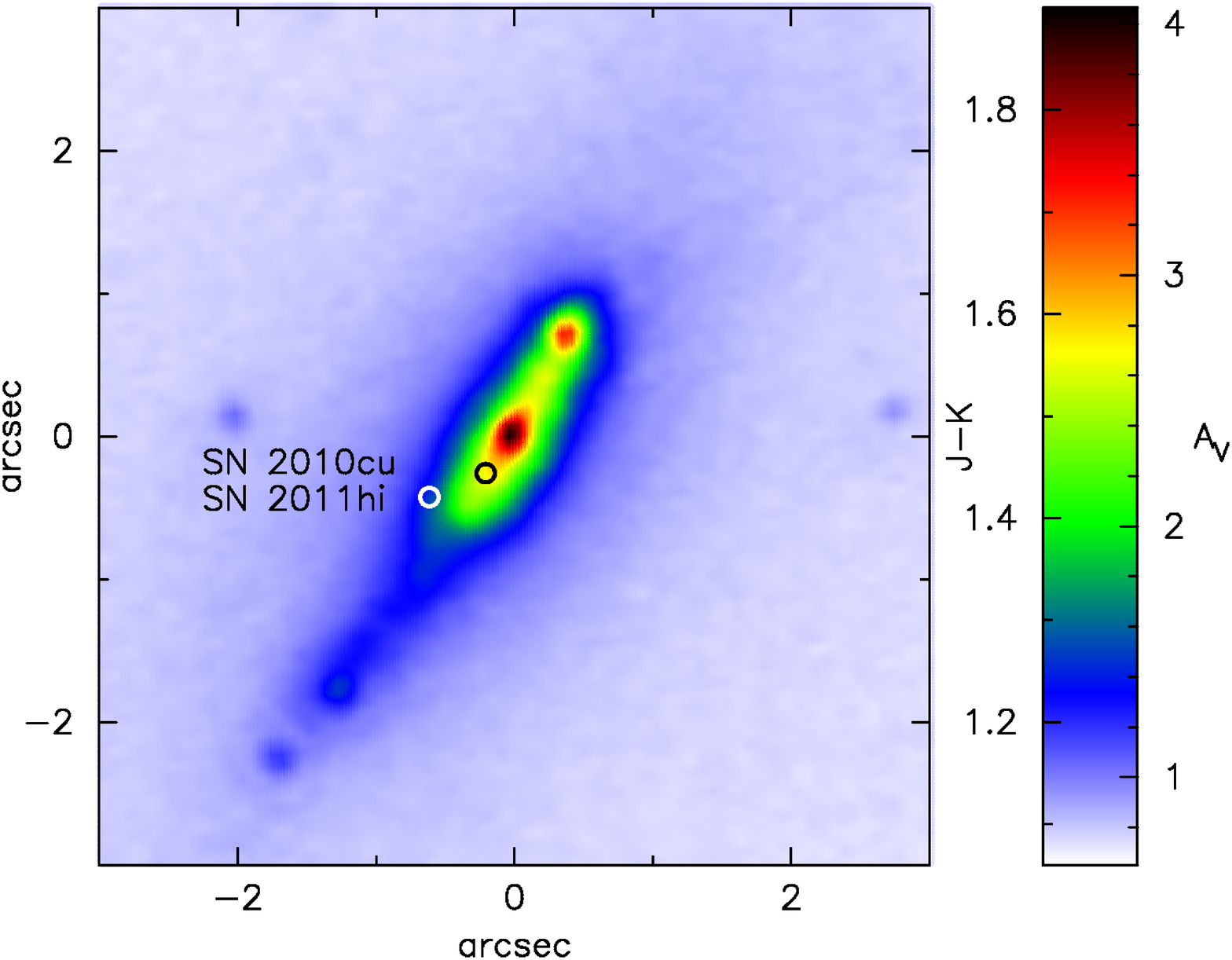}
\caption{6\arcsec$\times$6\arcsec \textit{J-K} \textit{HST}/NICMOS colour map of the central regions of IC 883 in Vega magnitudes converted to extinction adopting the Calzetti extinction law. North is up and East is to the left.}
\end{figure}

As part of a future effort to estimate CCSN rates in our sample of LIRGs, detection efficiencies for the Gemini images were derived by simulating artificial SNe in the images. For this the images were divided into multiple contour regions based on the surface brightness, a reference frame was subtracted from the images using ISIS 2.2 and the subtracted frames were run through SExtractor \citep{bertin96}. Aperture photometry of the detected objects was compared with the standard deviation calculated for a random grid of aperture sums within each contour region. This procedure was repeated for different simulation magnitudes and the 3$\times$ standard deviation detection efficiency for each magnitude and contour area was calculated. Using s-curve \citep{dahlen08} fitting to the results we derive a preliminary 50\% detection efficiency for 18.3~mag sources in our \textit{K}-band Gemini images in the innermost circumnuclear region in which both \object{SNe 2010cu} and \object{2011hi} were discovered. These initial results indicate that our search was sensitive enough to discover SNe $\sim1-1.5$~mag fainter than \object{SNe 2010cu} and \object{2011hi} in the central regions of \object{IC 883} (and LIRGs with similar properties). This corresponds to $\sim10-15$~mag higher extinctions in $A_{V}$. However, our SN detection sensitivity drops significantly in the nuclear $<$100~pc regions \citep[see also][]{romero-canizales11a} with only very high resolution VLBI observations being able to detect highly obscured SNe therein \citep[e.g.,][]{perez-torres09}.

\section{Discussion}

For \object{SN 2010cu} we obtained satisfactory fits with both the 'slowly declining' and Type IIP templates whereas the light curves and colours of the 'ordinary' template fit are not consistent with the SN, when the \textit{J}-band upper limit is taken into account. The absolute brightness and the discovery epoch of the SN at day 62 given by the Type IIP fit is also very reasonable given the fact that the galaxy was observed for the first time in over 8 months. In contrast in the 'slowly declining' template fit the SN would have to be underluminous by more than 2~mag compared to the template and discovered close to the maximum. However, based on only three epochs of photometry we cannot definitively determine the type of \object{SN 2010cu} between 'slowly declining' and Type IIP. Regardless of the template used they both yield very low host galaxy extinction estimates $A_{V}\sim0-1$~mag.

For \object{SN 2011hi} we do not find the 'ordinary' template fit very likely, even though it gives the best reduced $\chi^{2}$ value, as it would require the SN to be discovered on day -38, thus having an unrealistically long rise time for an ordinary CCSN. We note that in \citet{kankare08} we found a $K$-band rise time of $>53$~days for \object{SN 2008cs} with the 'slowly declining' template providing the best fit to the data. However, CCSNe interacting with their CSM have shown delayed rise to their near-IR peak compared to their optical discovery, e.g., SN 2006jc \citep{mattila08}. The 'slowly declining' and the Type IIP templates both give reasonable fits and extinction estimates $A_{V}\sim0-7$~mag. In the Type IIP comparison the SN is expected to be 1.65~mag brighter than the canonical \object{SN 1999em}, however, Type IIP SNe are known to show intrinsic variations of several magnitudes in their maximum brightness. For example \object{SN 2009kf} \citep{botticella10} was roughly 2~mag brighter than a prototypical Type IIP. Adopting the 'slowly declining' template \object{SN 2011hi} would have been caught in the beginning of the near-IR excess phase, being intrinsically slightly brighter than the template. As only two epochs of photometry are available both template fits seem reasonable. 

Radio observations of \object{SNe 2010cu} and \object{2011hi} with the electronic European VLBI network (e-EVN) at 5~GHz on 2011 March 23.1 UT and the electronic Multi-Element Remotely Linked Interferometer Network (e-MERLIN) at 6.9~GHz resulted in upper limits for the radio emission from both SNe \citep[see][for details]{canizales11,romero-canizales11b}, which do not allow classifying them. In \citet{romero-canizales11b} an average SF rate of 185~M$_{\sun}$~yr$^{-1}$ for \object{IC 883} over the duration of the starburst was also derived based on the IR spectral energy distribution predicting a CCSN rate of 1.1~yr$^{-1}$, consistent with the discovery of two SNe within two years.

As shown earlier, the $J-K$ colour map extinction of $\sim3$~mag for the \object{SN 2010cu} line-of-sight through the whole galaxy is more of a lower limit for the total line-of-sight extinction. This gives freedom for any template fitted extinction estimate to be possible. However, the template fits for \object{SN 2010cu} give extinctions well below this limit. For \object{SN 2011hi} only the 'slowly declining' template fit with negligible host galaxy extinction is found to be consistent with an estimated maximum line-of-sight extinction of $A_{V}\sim1.5$~mag from the extinction map. However, higher localized extinction values, beyond the resolution of the extinction map, could be possible. 

\section{Conclusions}

The adaptive optics assisted discovery of two consecutive SNe located at $<400$~pc from the LIRG nucleus demonstrates the importance of high spatial resolution in searches for CCSNe in highly crowded starburst regions in LIRGs. We note that of the five SNe discovered using high resolution near-IR observations in our sample of LIRGs (see Sect.~1), four are within 1~kpc from the host nucleus. 

Based on the near-IR light curve fitting to templates we find \object{SN 2010cu} to be consistent with either a Type IIP or a near-IR bright Type IIn/L SN with a host galaxy extinction of $A_{V}\lesssim1$~mag. Taking into account also the extinction map \object{SN 2011hi} is found to be most consistent with a Type IIn/L SN during the near-IR excess phase with host galaxy extinction of $A_{V}\sim0$~mag. A Type Ib/c classification was found unlikely for both the SNe. Future near-IR and radio follow-up observations of \object{SN 2011hi} will be able to better constrain the nature and line-of-sight extinction of this potentially very interesting event  

The surprisingly low derived line-of-sight extinctions appear to be in conflict with the expectation of the occurrence of highly obscured CCSNe in the nuclear region of LIRGs. For example, in \citet{kankare08} $A_{V}=16$~mag of host galaxy extinction was derived with near-IR template light curve fitting for \object{SN 2008cs}, located at a projected distance of 1.5~kpc from the host LIRG nucleus. Based on this it seems more likely that \object{SNe 2010cu} and \object{2011hi} are located in the outer parts of the LIRG system which we see projected close to the LIRG nucleus. Additionally the majority of LIRGs at the high IR luminosity end are complex interacting systems and their dust extinction distributions can be very complex compared to normal spiral galaxies with symmetric exponential disks, see e.g., \citet{alonso-herrero06} and \citet{vaisanen08}. Despite the low extinctions such SNe occurring close to the LIRG nuclei remain undiscovered by the current SN searches and will therefore contribute to the 'SN rate problem' recently suggested by \citet{horiuchi11}. 

\acknowledgements
We thank the anonymous referee for useful comments and suggestions. EK acknowledges support from the Finnish Academy of Science and Letters (Vilho, Yrj\"{o} and Kalle V\"{a}is\"{a}l\"{a} Foundation). SM and EK acknowledge the support from the Academy of Finland (project: 8120503). AAH and LC acknowledge support from the Spanish Plan Nacional de Astronom\'{\i}a y Astrof\'{\i}sica under grant AYA2010-21161-C02-1. AA, MAP-T. and CR-C acknowledge financial support from the Spanish MICINN trough grant AYA2009-13036-CO2-01, co-funded with FEDER funds. JK acknowledge the support from the Academy of Finland (project: 2600021612)

\end{document}